\documentclass[a4paper]{article}
\usepackage{INTERSPEECH2020}
\usepackage{graphicx}
\usepackage{amssymb,amsmath,bm}
\usepackage{textcomp}
\usepackage{multirow}
\usepackage{caption}
\usepackage{subcaption}
\usepackage{verbatim}
\usepackage{soul}
\usepackage[table]{xcolor}

\title{On Bottleneck Features for Text-Dependent Speaker Verification Using X-vectors}
\name{Achintya Kumar Sarkar$^{1}$ and Zheng-Hua Tan$^{2}$}

\address{
  $^1$Indian Institute of Information Technology (IIIT), Sri City, India\\
  $^2$Department of Electronic Systems, Aalborg University, Denmark}
\email{sarkar.achintya@iiits.in, zt@es.aau.dk}

\begin{document}

  \maketitle
  
  \begin{abstract}
Applying x-vectors for speaker verification has recently attracted great interest, with the focus being on text-independent speaker verification. In this paper, we study x-vectors for text-dependent speaker verification (TD-SV), which remains unexplored. We further investigate the impact of the different bottleneck (BN) features on the performance of x-vectors, including the recently-introduced time-contrastive-learning (TCL) BN features and phone-discriminant BN features. TCL is a weakly supervised learning approach that constructs training data by uniformly partitioning each utterance into a predefined number of segments and then assigning each segment a class label depending on their position in the utterance. We also compare TD-SV performance for different modeling techniques, including the Gaussian  mixture  models-universal  background  model (GMM-UBM), i-vector, and x-vector. 
Experiments are conducted on the RedDots 2016 challenge database. It is found that the type of features has a marginal impact on the performance of x-vectors with the TCL BN feature achieving the lowest equal error rate, while the impact of features is significant for i-vector and GMM-UBM. The fusion of x-vector and i-vector systems gives a large gain in performance. The GMM-UBM technique shows its advantage for TD-SV using short utterances. 
\end{abstract}

\noindent{\bf Index Terms}: TCL bottleneck feature, i-vector, x-vector, GMM-UBM, text-dependent speaker verification

\section{Introduction}
Speaker verification (SV) is defined as the task of verifying a person using their speech/voice signal \cite{Bimbot04}. SV is an non-invasive bio-metric authentication method  with many real-world applications, e.g. alternative to passwords, airport security, banking transactions and home automation. SV methods can be broadly categorized into text-independent (TI) and text-dependent (TD). In TI-SV, speakers are free to speak any text or sentence during the enrollment and test phases. In TD-SV, on the other hand, speakers are constrained to speak predefined text during both enrollment/training and test phrases. Therefore, TD-SV maintains the matched phonetic context between the  training and test phases and yields a lower error rate than TI-SV, in particular for SV using short speech utterances of 
few seconds long, which is ideal for real-life applications.

In the model domain, GMM-UBM with maximum-a-posteriori (MAP) adaptation \cite{reynold00} and i-vector \cite{Deka_ieee2011} techniques are popular. It is well known that GMM-UBM outperforms the i-vector technique  for speaker verification using short utterances \cite{Delgado2016Asru}. 
With the recent progress of {deep neural networks} (DNNs), a new speaker embedding technique has been introduced for speaker recognition, which is called \emph{ x-vector} \cite{DBLP:conf/icassp/SnyderGSPK18} and has attracted much attention. In this method, a DNN is trained to model the variable
length speech segments in the initial few layers and discriminate the speakers at the output layer. Then the output of a particular DNN hidden layer for a given speech signal is used as a vectorized representation (i.e. x-vector) for the particular speech signal. During enrollment, target speakers are represented by x-vectors and during the test, the x-vector of a test utterance is scored against the claimant specific x-vector with probabilistic linear discriminate analysis (PLDA).  The effectiveness of x-vector for TI-SV using cepstral feature and senone {discriminant} 
BN features  can be found in \cite{DBLP:conf/icassp/SnyderGSPK18, Rahman2018}, and it is observed in \cite{Rahman2018}  that phonetic information is valuable for the speaker embedding using DNN.  To the best of our knowledge, the study of using x-vectors for SV has been focused on the text-independent setting only, and there is a missing comparison between x-vector with GMM-UBM in general in the literature.

On the other hand, many techniques have been proposed in the literature to improve the performance of speaker verification 
\cite{reynold00,Deka_ieee2011, DBLP:conf/icassp/SnyderGSPK18,kinnunen2010overview, Wang2019,Garcia-Romero2019,9053871} 
They can be grouped into two broad categories: feature domain and model domain.  
Feature domain methods include Mel-frequency cepstral coefficients (MFCCs) \cite{Davis80}, perceptual linear predictive (PLPs) \cite{Hermansky90};  and DNNs based bottleneck (BN) features \cite{DBLP:journals/taslp/SarkarTTSG19,yu2017adversarial,DBLP:journals/spl/SarkarDLB14}. When extracting BN features, it is common to feed cepstral features to a DNN to discriminate speakers \cite{Yuan2015}, senones \cite{McLaren2015}, a combination of them \cite{Yuan2015}, or tri-phone state \cite{Yuan2015} at the output layer. Afterward, the output of a particular DNN hidden layer called \emph{ deep feature} is projected onto a low dimensional space via principal component analysis (PCA) to obtain the BN feature. 
Recently, time-contrastive-learning (TCL) and phone discriminate DNN BN features have been introduced for TD-SV in \cite{DBLP:journals/taslp/SarkarTTSG19}. In TCL, text dependent pass-phrase utterances (excluding the evaluation set) are split into a predefined number of segments. The number of segments is equal to the number of  classes. The frames within a particular segment are assigned the same class label. Then a DNN is trained to discriminate the classes.
In case of the phone BN feature, a DNN is trained to discriminate the phones at the output layer. 
It is shown in \cite{DBLP:journals/taslp/SarkarTTSG19} that TCL and phone BN features give lower error rates than other existing BNs with both GMM-UBM and i-vector frameworks. However, the performance/behaviour of these features on the emerging x-vector paradigm has not been investigated. 

\begin{figure*}[t]
\centering\includegraphics[height=4.8cm,width=17.0cm]{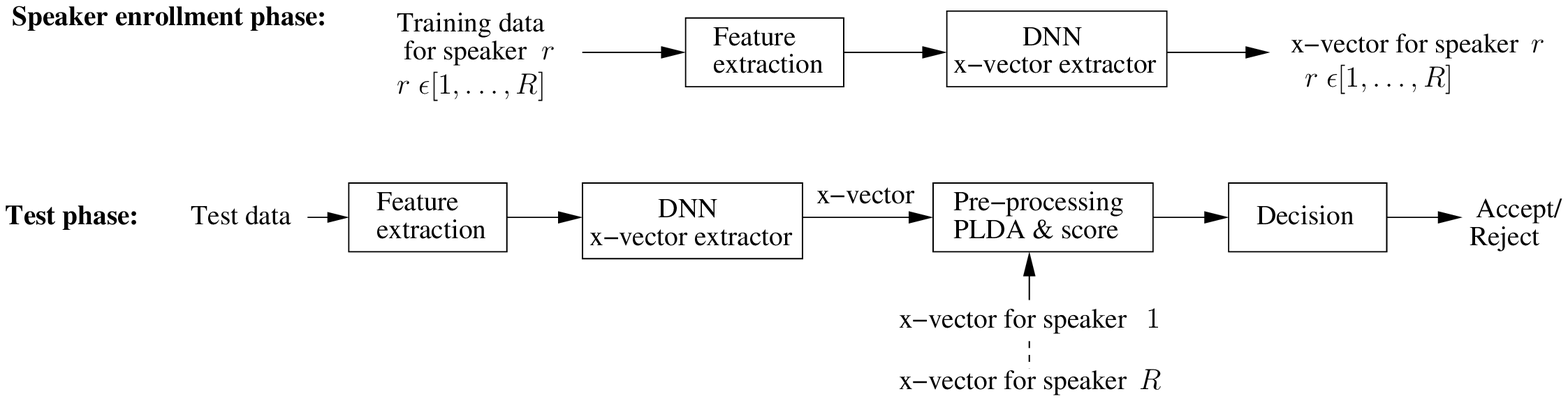}
\caption{Diagram of an x-vector system for text-dependent speaker verification.}
\label{fig:xvector_con}
\vspace*{-0.3cm}
\end{figure*}

The two sets of observations above motivate us to study the use of x-vector for TD-SV, in comparison with i-vector and GMM-UBM, and the impact of different bottleneck features on its performance. 
We conduct experiments on the RedDots 2016 challenge database \cite{RedDots}.
We show  the type of features has a marginal impact on the performance of x-vector with the TCL BN feature achieving the lowest equal error rate. On the other hand, the performance gap of using different BN features on i-vector is significant.  The fusion of x-vector and i-vector systems gives a large gain in performance.  The GMM-UBM technique shows its advantage for TD-SV using short utterances. 


The paper is organized as follows: Sections \ref{sec:mod_technique} \& \ref{sec:BN_feature} describe the modeling and BN extraction techniques for TD-SV, respectively. 
  The experimental setup is presented in Section \ref{sec:expsetup}. Section \ref{sec:resdis} presents the results and discussion. Finally, the paper is concluded in Section \ref{sec:con}.

\section{Modeling techniques} 
\label{sec:mod_technique}
In this section, we briefly describe the different modeling techniques, which are commonly used in speaker verification.
\subsection{GMM-UBM}
\label{sec:gmm-ubm}
In this approach \cite{reynold00}, a GMM based universal background model is trained using data of many non-target speakers. Then speaker models are derived from the GMM-UBM $\lambda_{ubm}$ with MAP adaptation. During test, a test utterance $\mathbf{X}=\{\mathbf{x_1},\mathbf{x_2}, \ldots, \mathbf{x_T}\}$ is scored against the claimant $\lambda_{tar}$ and GMM-UBM models. Finally, log likelihood ratio is calculated as
\begin{equation}
LLR(\mathbf{X}) = \frac{1}{T} \sum_{t=1}^{T}\{\log \;p(\mathbf{x_t}|\lambda_{tar}) - \log\; p(\mathbf{x_t}|\lambda_{ubm})\}
\end{equation}

\subsection{i-vector }
\label{sec:i-vect}
In this approach \cite{Deka_ieee2011}, a speech signal is represented using a low-dimensional vector {called i-vector, which} is obtained by projecting the signal onto a low dimensional subspace (called total variability (T) space) of a speaker independent GMM-UBM super-vector, where speaker and channel information is assumed to be dense. For a given speech signal of a speaker,  the speaker and channel dependent GMM super-vector $M$ can be expresses as 

\begin{equation}
 M = m + Tw 
\end{equation}
where  m denotes the speaker-independent GMM super-vector. and {$w$ is called an i-vector}.
During the enrollment phase, each target is represented by an average
i-vector computed over his/her training utterance-wise (or
speech session-wise) i-vectors. In the test phase, i-vector of a test utterance $w_t$  is scored against the  claimant specific i-vector $w_e$ (obtained during enrolment) with PLDA\cite{SenoussaouiInterspch2011}.

\begin{eqnarray}
score(w_e,w_t)=\log\frac{p(w_e,w_t|\theta_{tar})}{p(w_e,w_t|\theta_{non})} \label{eq:pldascore}
\end{eqnarray}

where $\theta_{tar}$ and $\theta_{non}$ indicate that both $w_e$ and $w_t$ are coming from a same or different speakers, respectively.

\subsection{x-vector}
\label{sec:x-vect}
In this method \cite{DBLP:conf/icassp/SnyderGSPK18}, a speech utterance is characterized by a vector that is obtained as the output of a hidden layer of a DNN, and the vector is called \emph{x-vector}.  The DNN is trained to model speech segments of variable lengths 
in the first several layers and embed the speakers at the last hidden layers. The loss function is the cross-entropy loss used to discriminate speakers at the output layer. Similarly to the i-vector system, speakers are represented by their average x-vectors computed over their training speech utterances in the enrollment phase. In the test phase, the x-vector of the test utterance is scored against the claimant specific x-vector with PLDA. Several studies using x-vector can be found in speaker verification with multi-conditional recordings \cite{8683760}, and in language recognition with triphone-states discriminant BN features trained using single or multiple languages \cite{SnyderOdys}.   More details about the x-vector technique can be found in \cite{DBLP:conf/icassp/SnyderGSPK18}.  Fig. \ref{fig:xvector_con} illustrates TD-SV using x-vectors.

\section{Bottleneck features}
\label{sec:BN_feature}
In this section, we briefly present the various bottleneck feature extraction methods used for TD-SV in this work.
\subsection{Speaker discriminant BN (spkr-BN)}
A DNN is trained to optimize a cross-entropy based objective function for discriminating speakers at the output layer \cite{Yuan2015}. The cross-entropy function can be defined as
\begin{eqnarray}
L (\theta) = - \frac{1}{T} \sum_{t=1}^{T} y_t\log p(\mathbf{x_t}, \theta)
\end{eqnarray}
where $L$, $\theta$, $y_t$, $\mathbf{x_t}$  and $p(.)$ denote the loss, parameters of DNN, the class label of the $t^{th}$ input feature vector and a posteriori output at the DNN output layer, respectively. The output of a particular hidden layer for a given speech segment is projected onto the low dimensional space to get the spkr-BN feature using PCA.

\subsection{Speaker+pass-phrase discriminant (spkr+phrase-BN)}
This system is analogous to \emph{speaker discriminant} BN. A DNN is trained to optimize two cross-entropy based objective functions simultaneously \cite{Yuan2015}: one for discriminating speakers $L_1(\theta)$ and the other for pass-phrase $L_2(\theta)$ defined on two different sets of output nodes 
\begin{equation}
L (\theta) =  \gamma L_1(\theta) + (1- \gamma) L_2(\theta)
\end{equation}
where $ 0 \leq \gamma \leq 1$. In our case, equally important is given to the two functions.

\subsection{Phone discriminant (PHN-BN)}
This system is similar to the \emph{spkr-BN}. The only difference is that phones are discriminated at the output layer of DNNs \cite{DBLP:journals/taslp/SarkarTTSG19}. The phone labels are obtained by using automatic speech recognition (ASR) systems. 
Three ASR systems are considered for generating the transcription of speech signals, yielding three different systems: a) {\bf PHN-BN1}: the phoneme recognizer is based on \cite{Schwarz},  
b)  {\bf PHN-BN2}: the phone recognizer is based on an end-to-end segmental phoneme recognizer \cite{tang2017endtoend},  
and c) {\bf PHN-BN3}:  this system considers forced-alignment for phone recognition, which is based on the end-to-end segmental model the same as in \emph{PHN-BN2}. 
Frames detected as \emph{sil} and \emph{pause}   are discarded before feeding feature vectors into respective DNNs.
More details analysis can be found in \cite{DBLP:journals/taslp/SarkarTTSG19}.


\subsection{Time-contrastive learning (TCL) }
\label{sec:tcl-bn}
The objective behind this feature is to capture the temporal information available from the  speech utterances in unsupervised manner i.e. without any ASR or manual transcriptions \cite{DBLP:journals/taslp/SarkarTTSG19}.
There are two settings for the method. In the first one, training data of the DNN are first randomized and then split into chunks of M frames with $M=6$ in this work. For the $c$ number of classes in TCL, $c$ segments are taken at a time and the frames within the $n^{th}$ segments are assigned class label $n$ {as}
\footnotesize{
\begin{equation}
\underbrace{(x_1, ..., x_M)}_\text{class $1$}, \ldots, \underbrace{(x_{iM+1}, ...,\\ x_{iM+M})}_\text{$\ldots$}, \ldots, \underbrace{(x_{(c-1)M+1}, ..., x_{cM})}_\text{class $c$}
\end{equation}
}

This is called \emph{stream-wise TCL (sTCL)}.  Similarly to \emph{spkr-BN} and \emph{PHN-BNs}, a DNN is trained to discriminate the classes at the output layer of DNN with a cross-entropy function {and afterward, BN features are extracted}. 

In the second setting, each utterance 
is split uniformly into $c$ segments, corresponding to $c$ classes, and all frames within one segment are assigned the same class label, which we call \emph{utterance-wise TCL (uTCL)}.  Afterward, a DNN is trained similarly to sTCL, and BN features are extracted. {In this study, we consider the value of $c=10$  as per \cite{DBLP:journals/taslp/SarkarTTSG19}.  }

\section{Experimental setup}
\label{sec:expsetup}
Experiments are conducted on the \emph{m-part-01} task (for male speakers) of the RedDots database as per protocol  \cite{RedDots}. There are $320$ target models that are trained by three utterances each. Each utterance is approximately $2$ seconds duration on average. There are four different types of trials for system evaluation as detailed in Table \ref{table:no_trial}.  More detail about the trials and plan can be found in \cite{RedDots}.

MFCC feature vectors of $57$ dimensions consisting of $19$ static and their $ \Delta, \Delta\Delta$  are extracted from speech signals using a $20 ms$ hamming window and a $10ms$ frame shift. An energy-based voice activity detector is used to discard the less energized frames. The selected frames are normalized to fit zero mean and unit variance at the utterance level.
A GMM-UBM with $512$ mixtures having diagonal co-variance matrices is trained using $6300$ speech files from the TIMIT database consisting of $438$ males and $192$ females.  This data set is also used for training PCA to get low dimensional BN features. In MAP adaptation, $3$ iterations and value of relevance factor $10$ are used.

$T$-space in the i-vector system is trained using $72764$ utterances covering $27$ pass-phrases from the RSR2015 database consisting of $157$ male and $143$ female speakers, while excluding the pass-phrases common/overlapping with the RedDots database.  
This data set is also used for the PLDA, x-vector and DNNs training. These numbers result in $300$ and $327$ nodes at the output layer of DNNs for \emph{spkr-BN} and \emph{spkr+phrase-BN} systems, respectively.  

Kaldi toolkits \cite{xvector_recipe} is used for implementing the x-vector technique, i.e. the speaker embedding part. The number of DNN layers, activation functions, the number of neurons per layer and other  parameters are considered as per \cite{xvector_recipe}. $400$ dimensional x-vectors are extracted to align with the dimension of i-vector based systems. Kaldi truncates the training data into chunks, 
and due to short utterances used for training in this work, we set  minimum and maximum chunk sizes as $50$ and $150$, respectively, in contrast to the default $200$ and $400$ frames.
Speaker having the same pass-phrase utterances are considered as a separate speaker/class and gives  $\approx 7075$ speakers/classes in embedding.

CNTK toolkit \cite{YuCNTK2014} is used for implementing the bottleneck feature extraction {with the following settings: variable batch sizes from $256$ to $1024$,  variable learning rates from $0.8$ 
to $0.08$, and $25$ training epochs, 
as per the default parameters settings}. Seven layers feed forward networks with a sigmoid activation function is used. Each hidden layer consists of $1024$ neurons.  For BN feature extraction, the fourth hidden layer for \emph{spkr}, \emph{spkr+phase} and the second hidden layer of DNNs  for \emph{PHN} and \emph{TCLs}  are projected onto the low dimensional space as per \cite{DBLP:journals/taslp/SarkarTTSG19}.

\begin{table}[t!]
\begin{center}
\caption{Number of different trials available in RedDots on \emph{m\_part\_01} task.}
\begin{tabular}{|l|l|l|l|}\cline{1-4}
\# of  & \multicolumn{3}{c|}{\# of trials in Non-target type} \\ 
  Genuine            & Target  & Imposter & Imposter \\
  trials             &-wrong   &  -correct &  -wrong    \\   \hline
       3242       & 29178       & 120086            & 1080774  \\ \hline
\end{tabular}
\label{table:no_trial}
\end{center}
\vspace*{-0.7cm}
\end{table}

 In PLDA, the utterances of the same pass-phrase from a particular speaker are treated as an individual speaker. It gives $8100$ classes (4239 males and 3861 females) in PLDA. Speaker and channel factors are kept full  in PLDA, i.e. equal to the dimension of i-vector ($400$), x-vector ($400$) and vector-fusion ($800$, where i-vector is concatenated with x-vector per utterance-wise) in the respective systems. Before PLDA, i-vector and x -vectors are normalized with \emph{spherical normalization} of $2$ iterations  \cite{Pierre-interspeech2012}.   System performance is measured in terms of equal error rate (EER) and Minimum detection cost function (MinDCF) as per 2008 SRE \cite{SRE08}.

\begin{table*}[ht!]
\caption{Performance comparison across TD-SV systems using different features and modeling techniques, including GMM-UBM, i-vector, x -vector , and fusion of i-vector and x-vector. Bold font indicates the lowest EER among all systems for a particular feature. The experiments are on the m-part-01 task of the RedDots database. (i,x) denote the i-vector and x- vector systems, respectively. Data-driven is not applied in systems. In \cite{DBLP:journals/taslp/SarkarTTSG19}, with data-driven uTCL achieves the lowest EER among all systems.} 
\begin{center}
\begin{tabular}{|l|lllccc|c|}\cline{1-8}
& Feature & \# of classes & Method & \multicolumn{3}{c|}{Non-target type [\%EER/(MinDCF$\times$ 100)]} & Average \\  
&         & in BN        &    & Target-wrong& Impostor-correct  & Impostor-wrong     & (EER/MinDCF) \\  \hline 
& MFCC    &         & GMM-UBM  &5.12/2.17  & 3.33/1.40  & 1.14/0.47 & 3.19/1.35    \\  
\parbox[t]{2mm}{\multirow{15}{*}{\rotatebox[origin=c]{90}{Existing}}}&         &         &i-vector & 6.96/3.23    & 4.82/2.03    & 1.63/0.61      & 4.47/1.96 \\ 
&        &         &x-vector  & 4.28/1.89         & 5.36/2.37    & 1.07/0.38      & 3.57/1.55 \\
&        &         &          &                   &              &                &           \\
&        &          & score-fusion (i,x) & 3.96/1.79     & 3.94/1.70    & 0.67/0.25      &  2.86/1.25 \\
&        &          & vector-fusion (i,x) & 3.51/1.51     &3.95/1.70      & 0.80/0.26      & {\bf 2.75}/1.16 \\    \cline{2-8}
&SPKR-BN  & 300     &GMM-UBM & 4.59/1.65    & 3.05/1.35    &1.11/0.38    & 2.91/1.13\\
&         &         &i-vector  & 7.19/3.02   & 6.06/2.34    & 2.11/0.85      & 5.15/2.07 \\
&        &          & x-vector & 3.98/1.82   & 5.53/2.51    & 0.94/0.41      & 3.48/1.58 \\
&        &         &          &                   &              &                &           \\
&        &          & score-fusion (i,x) & 3.73/1.68   & 4.21/1.78   & 0.67/0.30      & {\bf 2.87}/1.25 \\
&        &          & vector-fusion (i,x)  & 3.84/1.55   & 4.47/1.86   & 0.86/0.35      & 3.01/1.26  \\ \cline{2-8}
&SPKR+    &          & GMM-UBM  &4.53/1.64    &3.07/1.34     &1.17/0.38    & 2.92/1.12 \\
&phrase-BN& 327     &i-vector  & 7.27/3.01    & 6.07/2.34    & 2.11/0.85     & 5.15/2.02 \\
&         &         &x-vector  & 4.10/1.82    & 5.15/2.34    & 1.07/0.42      &  3.44/1.53  \\ 
&        &         &          &                   &              &                &           \\
&         &         & score-fusion (i,x)  & 3.70/1.66    & 3.88/1.66    & 0.72/0.26      & {\bf 2.77}/1.19\\
&         &         & vector-fusion (i,x)  & 3.48/1.55    & 4.42/1.86    & 0.93/0.35  & 3.07/1.26 \\ \hline 
& PHN-BN1 & 38      & GMM-UBM  & 2.31/0.71    & 3.14/1.29    & 0.61/0.20   & {\bf 2.02}/0.73 \\
\parbox[t]{2mm}{\multirow{15}{*}{\rotatebox[origin=c]{90}{Phone-discriminate }}}&         &         &i-vector  & 2.68/1.04    & 4.57/1.94    & 0.89/0.26     & 2.71/1.08 \\ 
&         &         & x-vector & 3.48/1.34    & 5.79/2.65    &  1.14/0.46    & 3.47/1.49 \\
&        &         &          &                   &              &                &           \\
&         &         & score fusion (i,x)   & 2.06/0.84    & 4.15/1.73    & 0.50/0.16     &  2.24/0.91 \\
&         &         & vector-fusion (i,x) & 1.88/0.76    & 4.25/1.77   & 0.61/0.18 & 2.24/0.90 \\ \cline{2-8} 
&PHN-BN2  & 47      & GMM-UBM    & 2.25/0.78    & 2.89/1.30    & 0.61/0.22   & {\bf 1.92}/0.77  \\
&        &          & i-vector   & 2.87/1.15    & 4.71/1.86    & 0.89/0.30      & 2.83/1.10  \\ 
&        &          & x-vector    & 3.05/1.22    & 6.04/2.68 & 1.07/0.42   & 3.39/1.44 \\     
&        &         &          &                   &              &                &           \\
&        &          &score fusion (i,x) & 1.94/0.77    & 4.28/1.77       & 0.55/0.18   & 2.26/0.91 \\   
&        &          &vector-fusion (i,x) & 1.96/0.74  & 4.19/1.75 & 0.64/0.17 &  2.26/0.88\\ \cline{2-8} 
&PHN-BN3 & 39       & GMM-UBM & 1.79/0.72    & 3.08/1.41    &  0.55/0.15  & {\bf 1.81}/0.76 \\ 
&        &          & i-vector  &  2.25/0.83    & 4.65/1.90      & 0.89/0.26  & 2.59/1.00   \\
&        &          & x-vector  & 3.30/1.34     & 6.02/2.50      & 1.01/0.47  & 3.44/1.44 \\
&        &         &          &                   &              &                &           \\
&        &          & score-fusion (i,x) &  1.71/0.74   & 4.10/1.73     & 0.50/0.17   &  2.11/0.88 \\
&        &          &vector-fusion (i,x) & 1.81/0.65  & 4.31/1.72 & 0.52/0.19 & 2.22/0.85 \\ \hline 
& sTCL-BN   & 10       & GMM-UBM & 4.42/1.61       &  3.08/1.32    & 1.12/0.38  & 2.88/1.10 \\  
&        &          & i-vector& 6.60/2.97    & 5.51/2.25    &  1.80/0.74     & 4.63/1.99 \\
\parbox[t]{2mm}{\multirow{8}{*}{\rotatebox[origin=c]{90}{Time-contrastive }}}&        &          & x-vector & 3.95/1.83    & 5.66/2.66    & 1.01/0.40      &  3.54/1.63\\
 &        &         &          &                   &              &                &           \\
&        &          & score-fusion (i,x) & 3.36/1.64   & 3.91/1.85    & 0.74/0.26      & {\bf 2.67}/1.25 \\ 
&        &          & vector-fusion (i,x) & 3.36/1.52   & 4.21/1.88    & 0.89/0.28      & 2.82/1.23 \\   \cline{2-8}
& uTCL-BN   &  10      & GMM-UBM & 1.88/0.65     &  3.14/1.44   & 0.64/0.19  & {\bf 1.89}/0.76 \\
 &       &          & i-vector & 2.74/0.97    & 5.27/2.08    & 0.95/0.32   &  2.99/1.12  \\
 &       &          &x-vector &2.63/1.03  &  6.27/2.96         & 1.14/0.41   & 3.34/1.47  \\
 &        &         &          &                   &              &                &           \\
 &       &          & score-fusion (i,x) &1.75/0.67  & 4.60/1.93   & 0.58/0.18   & 2.31/0.93\\  
 &       &          & vector-fusion (i,x) &1.63/0.66  & 4.90/1.88   & 0.63/0.19   & 2.35/0.91 \\ \hline
\end{tabular}\\

\end{center}
\label{table:Table2}
\vspace*{-0.5cm}
\end{table*}

\section{Results and Discussions}
\label{sec:resdis}

Table \ref{table:Table2} compares the TD-SV performance of various features combined with GMM-UBM, i-vector or x -vector as well as the fusion of i-vector and x-vector on the RedDots database (m-part-01 task). 
{In score fusion, scores of the different systems are combined with equal weights.}  The following observations can be deduced from the table.

First, there are marginal performance differences among different features under the x-vector framework, although the uTCL BN feature gives the lowest EER. This could be due to x-vector being trained to neutralize the differences in representation power across BN features. On the other hand, the impact of features is rather a signification on i-vector and GMM-UBM, for which PHN-BN and TCL-BN outperform the others with big margins. In most cases, all BN features across the modeling methods outperform MFCC. 

Second, GMM-UBM gives the lowest error rates for all features explored, when a single modelling technique is used. This indicates that GMM-UBM is better than x-vector and i- vector each alone for TD-SV using short utterances. 

Third, the x-vector yields lower error rates than the i-vector for all features but PHN and TCL BNs. Fusion of x vector system with i-vector significantly reduces the EERs compared to their standalone, indicating their complementary nature. 

Finally, it is interesting to notice that i-vector performs better than x-vector when a strong-performing/highly-discriminative BN is used, e.g. PHN-BNs and TCL-BNs. It is the opposite when MFCCs, SPK-BN, and SPK-phrase-BN are used. 

{It is worth to mention that score fusion of GMM-UBM, i-vector and x-vector systems does not lead to further performance improvement and hence the results are not shown in this paper. More complex fusion strategies will be explored as future work.}

\section{Conclusion}
\label{sec:con}
In this paper, we studied the use of x-vector and its combination with various bottleneck (BN) features for text-dependent speaker verification (TD-SV) using short utterances. We further compared the TD-SV performance of x-vector with Gaussian mixture models-universal background model (GMM-UBM) and i-vector methods.   Experiments lead to a set of interesting results. First, BN features have a marginal impact on the performance of x-vector, while they have a large impact for i-vector and UBM-GMM in favor of phone-discriminant and time-contrastive-learning BN features. The fusion of x-vector and i-vector largely boosts the performance, while GMM-UBM remains a favorable framework for TD-SV with short utterances. 

 \clearpage
 \bibliographystyle{IEEEbib}
 \bibliography{strings,References}
 \end{document}